\documentclass[12pt]{iopart}

\usepackage{graphicx}
\begin{document}

\title{Search for Conical Emission with Three-Particle Correlations}
\author{Claude A. Pruneau, for the STAR Collaboration}
\address{Physics and Astronomy Department, Wayne State University, Detroit, MI 48152 USA}
\ead{pruneau@physics.wayne.edu}
\begin{abstract}
We present preliminary STAR results on 3-particle azimuthal angle correlation studies in d+Au and Au+Au collisions at $\sqrt{s_{NN}}$ = 200 GeV.
 The studies are carried out at mid-rapidity between a trigger particle with 3 $\le p_{\perp} \le$ 4 GeV/c and two associated particles in 1 $\le p_{\perp} \le $ 2 GeV/c.
 A cumulant analysis reveals finite 3-particle azimuthal correlations, dominated by near and away side particle correlations consistent with jet production, and jet-flow correlations.
 We use a two-component model to remove underlying background correlations. 
This analysis indicates the presence of the conical emission signals in
central Au+Au collisions within the model assumptions about background
composition and normalization.
\end{abstract}


\section{Introduction}

The observation of a dip at $180^o$ in flow subtracted two-particle azimuthal correlations observed in 
 Au + Au collisions \cite{Star05,Ulery05} was suggested as an indicator of the production of away-side parton induced wake field or Mach 
cone \cite{Stoecker05}. The dip might however also result from large angle gluon radiation,  jets deflection by radial flow, or Cerenkov gluon radiation \cite{Vitev05}. 
 While discrimination of these production mechanisms is not possible with two-particle 
correlations, it might be achieved with three particle correlations.  

We present the results of {\bf two} parallel approaches, by  STAR, to search for conical emission in Au + Au collisions at $\sqrt{s_{NN}}$=200 GeV  using  
three particle cumulants \cite{Pruneau06} and a two-component model based subtraction method \cite{Ulery06}. 

The analyses use Au+Au datasets consisting of nearly 20 million minimum bias and 
22 million central-trigger collected during RHIC run 4. Correlations are measured between three charged particles in the range $| \eta | <1$
in terms of two relative azimuthal angles $\Delta \varphi _{ta}$ and  $\Delta \varphi _{tb}$.
 Particle "t",  selected in the range $3 \le p_ \bot   \le 4{\rm{ GeV/c}}$,
serves as a jet tag or trigger while particles "a" and "b", in the range $1 \le p_ \bot \le 2{\rm{ GeV/c}}$, 
probe the jet structure and presence of conical emission.  Mach cone emission shall lead to jacobian peaks at  $\Delta \varphi _{ta} = \pi \pm \theta_{M}$, $\Delta \varphi _{tb} = \pi \pm \theta_{M}$, 
with the Mach angle $\theta_{M}$ determined by the sound velocity in the produced medium.

Preliminary three particle analyses were presented at QM05 \cite{Ulery05}, and elsewhere\cite{Ulery05a}. 

\section{Three-Particle Cumulant Analysis}

The 3-cumulant, introduced in \cite{Pruneau06} and also discussed in \cite{Pruneau06a} is defined as
\begin{equation}
 C_3 (\Delta \varphi _{(ta)} ,\Delta \varphi _{(tb)} ) = \rho _3  - \rho _2^{(ta)} \rho _1 ^{(b)} 
 - \rho _2^{(tb)} \rho _1 ^{(a)}   - \rho _2 ^{(ab)} \rho _1 ^{(t)} + 2\rho _1 \rho _1 \rho _1 
\end{equation}
in terms of the measured 3-particle density (number of triplets/event) $\rho _3\equiv dN/d\Delta \varphi _{ta} d\Delta \varphi _{tb}$ and combinatorial terms $\rho _2 ^{(ij)} \rho _1 ^{(k)}$
 and $\rho_1\rho_1\rho_1$ calculated based on measured 2- and 1- particle densities. 
Figure 1 presents the 3-cumulants measured in  three ranges of  centrality.
The 3-cumulants feature prominent peaks at (0,0), ($\pi$,$\pi$), (0,$\pi$), and ($\pi$,0). 
One also notes the presence of weaker peaks positioned at regular intervals. These are strongest in 10-30\% collisions  where $v_2$ and $v_4$ flow coefficients are the largest, and
are qualitatively understood as arising from irreducible non-diagonal collective flow contributions of order $v_2v_2v_4$ \cite{Pruneau06}.  
Calculation of these contributions based on parameterizations of measured $v_2$ and $v_4$ \cite{YutinBai06} yields amplitudes compatible with those observed in Fig. 1. 
The near- and away-side structures are present at all centralities, including the most central collisions. While their presence can in part result from jet emission alone (with two 
associated particles either on the near or away-side), they may also result from the interplay of 
jet correlations with the event reaction plane and particle flow. Their interpretation thus requires further work.
Note finally the away-side structures are inconsistent with global momentum conservation effects expected to produce broad structures in three-particle correlations \cite{Borghini06}. 

\begin{figure}[htb]
\centerline{\includegraphics[height=3.8in,width=6.5in]{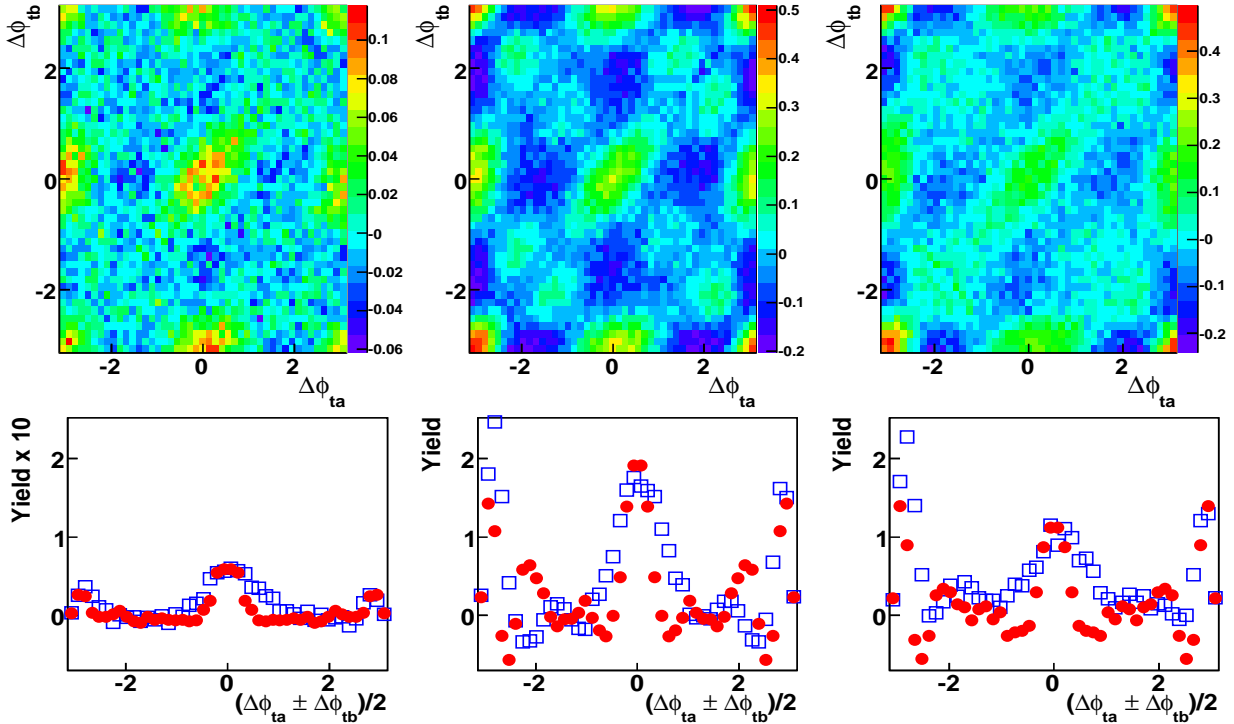}}
\caption[]{3-cumulants for (left) 80-50\% (scaled x 10) (center) 30-10\% and (right) 10-0\%, and projections along the main (blue open squares) and alternate diagonals (red circles). All results are preliminary.}
\label{fig1}
\end{figure}
Next we examine the 3-cumulant and their projections (bottom part of Fig 1.) to seek evidence of structures produced by conical emission \cite{Stoecker05,Pruneau06} along the diagonals 60-80$^o$ from the away-side direction. Projections axes are set to show away-side peak at 0 radian.
The projections exhibit structures consistent with $v_2 v_2 v_4$  terms but no clear evidence for conical emission. 
It is however conceivable that conical emission is masked by flow terms, or simply too weak to be visible in this analysis. 

\section{Two component Subtraction Method}
Preliminary results obtained with this analysis technique described in \cite{Ulery06} were already reported \cite{Ulery05,Ulery06a}.  
Current results are also presented in QM06 poster proceedings \cite{Ulery06b}. 
This method is based on a measurement of three-particle density normalized per trigger particle.  It includes explicit removal of $v_2^2$, 
$v_4^2$, $v_2v_2v_4$ terms, and carries combinatorial background term subtraction using a zero yield at 1 radian (ZYA1) hypothesis \cite{Ulery06}.
Fig. 2 presents the correlations obtained after flow and combinatorial background subtraction for $d+Au$, and selected $Au + Au$ centralities.  
The correlation obtained for 80-50\%  $Au + Au$ collisions is qualitatively similar in shape to that obtained in $d+Au$ as well as that measured with the cumulant method.
At 12-0\% centralities, the shape of the away-side structure is however substantially modified relative to the peripheral bin: 
two broad peaks are observed along the main diagonals at angles of $\pi \pm 1.45$ rad. One also observes off diagonal 
structures at ($\pi \pm 1.45$, $\pi \mp 1.45$). 
Diagonal projections shown in Fig. 2 (bottom) indicate these structures are statistically significant, and thus suggest evidence for conical 
emission in the most central bin shown, although the amplitude of the structures depends on the inclusion of a 
flow-jet correlation component, the amplitude of the $v_2$ and $v_4$ 
flow, and combinatorial terms normalization.   Systematic effects dominated by flow subtraction and background normalization are shown as (yellow) histogram band in Fig 2.
The position of the peaks angles, at $\pi \pm 1.45$, if arising from Mach cone emission, would suggest the sound velocity is rather modest in the 
produced medium \cite{Stoecker05}. Further work is in progress to fully understand these effects. 
\begin{figure}[htb]
\mbox{
\begin{minipage}{0.33\linewidth}\begin{center}\includegraphics[width=1.8in,height=2.in]{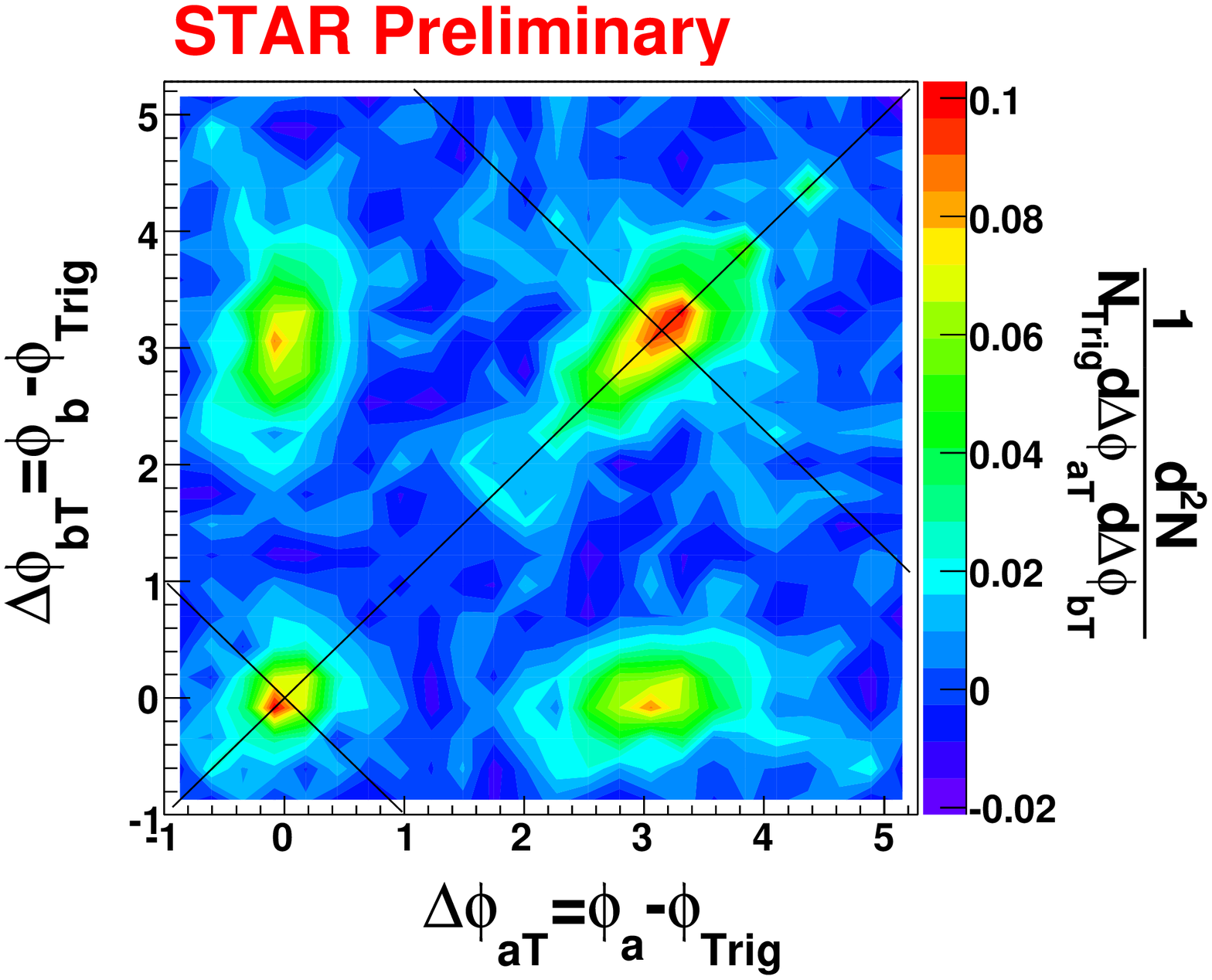} \end{center}\end{minipage}
\begin{minipage}{0.33\linewidth}\begin{center}\includegraphics[width=1.8in,height=2.in]{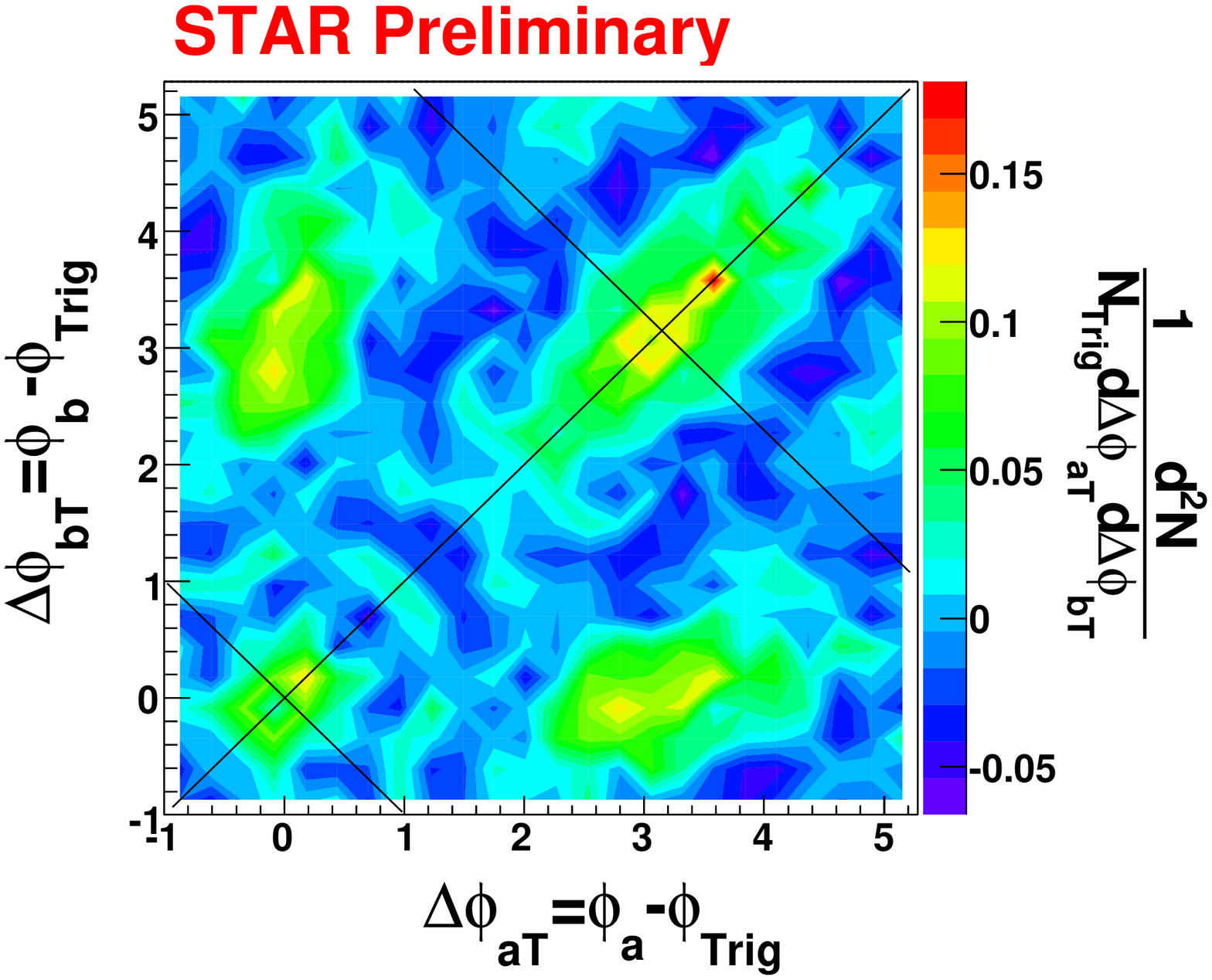} \end{center}\end{minipage}
\begin{minipage}{0.33\linewidth}\begin{center}\includegraphics[width=1.8in,height=2.in]{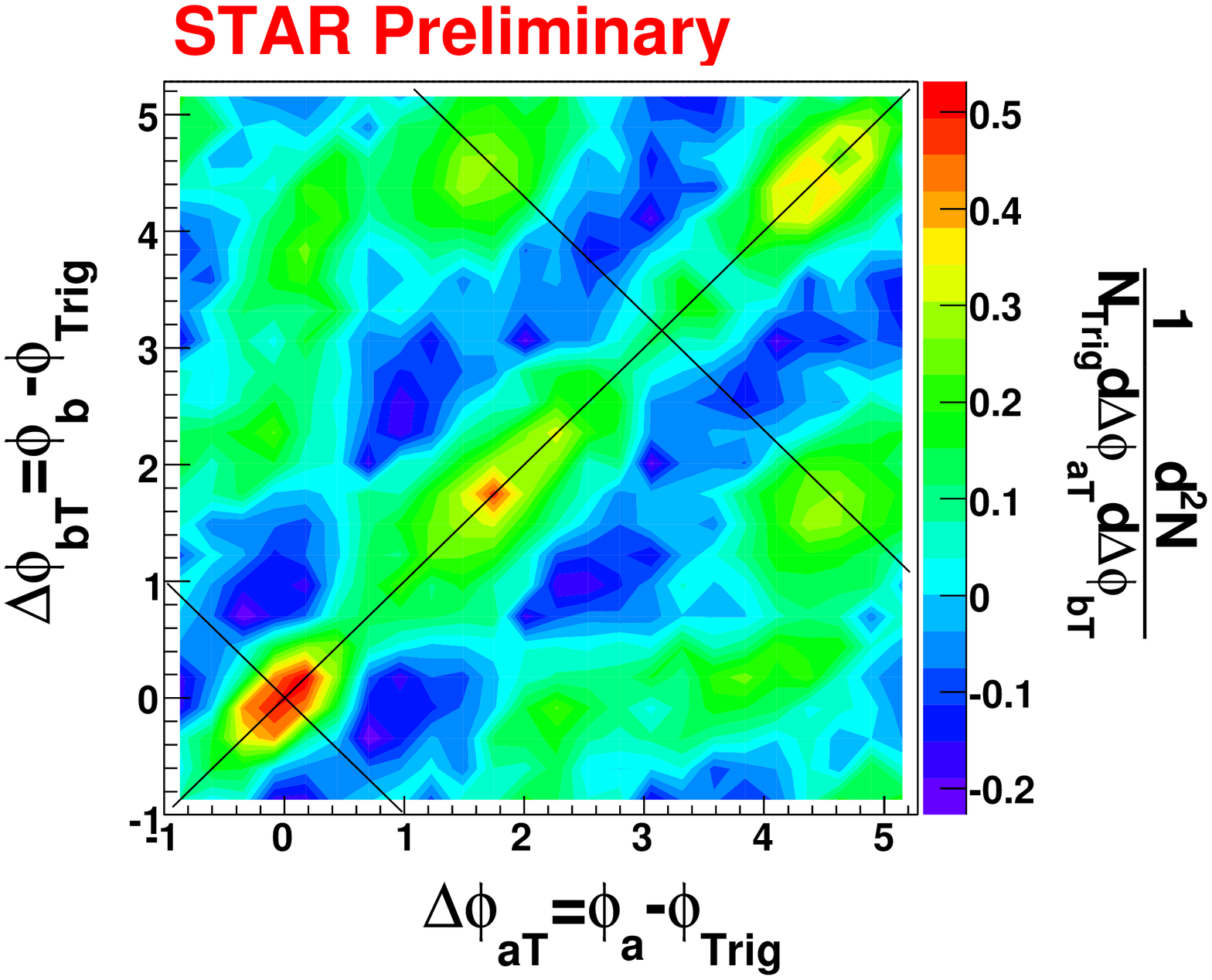} \end{center}\end{minipage}
}
\mbox{
\begin{minipage}{0.33\linewidth}\begin{center}\includegraphics[width=1.8in,height=2.in]{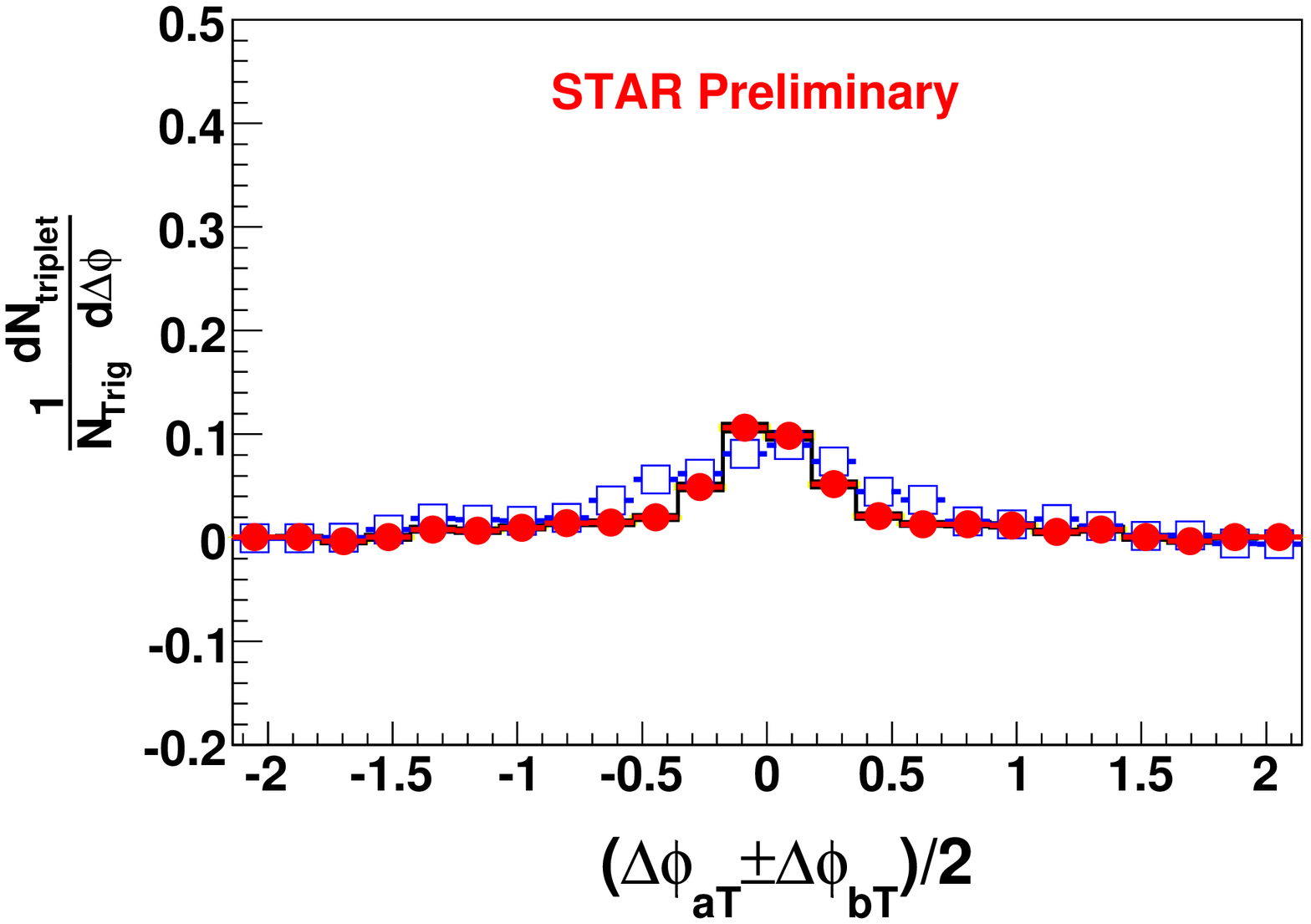} \end{center}\end{minipage}
\begin{minipage}{0.33\linewidth}\begin{center}\includegraphics[width=1.8in,height=2.in]{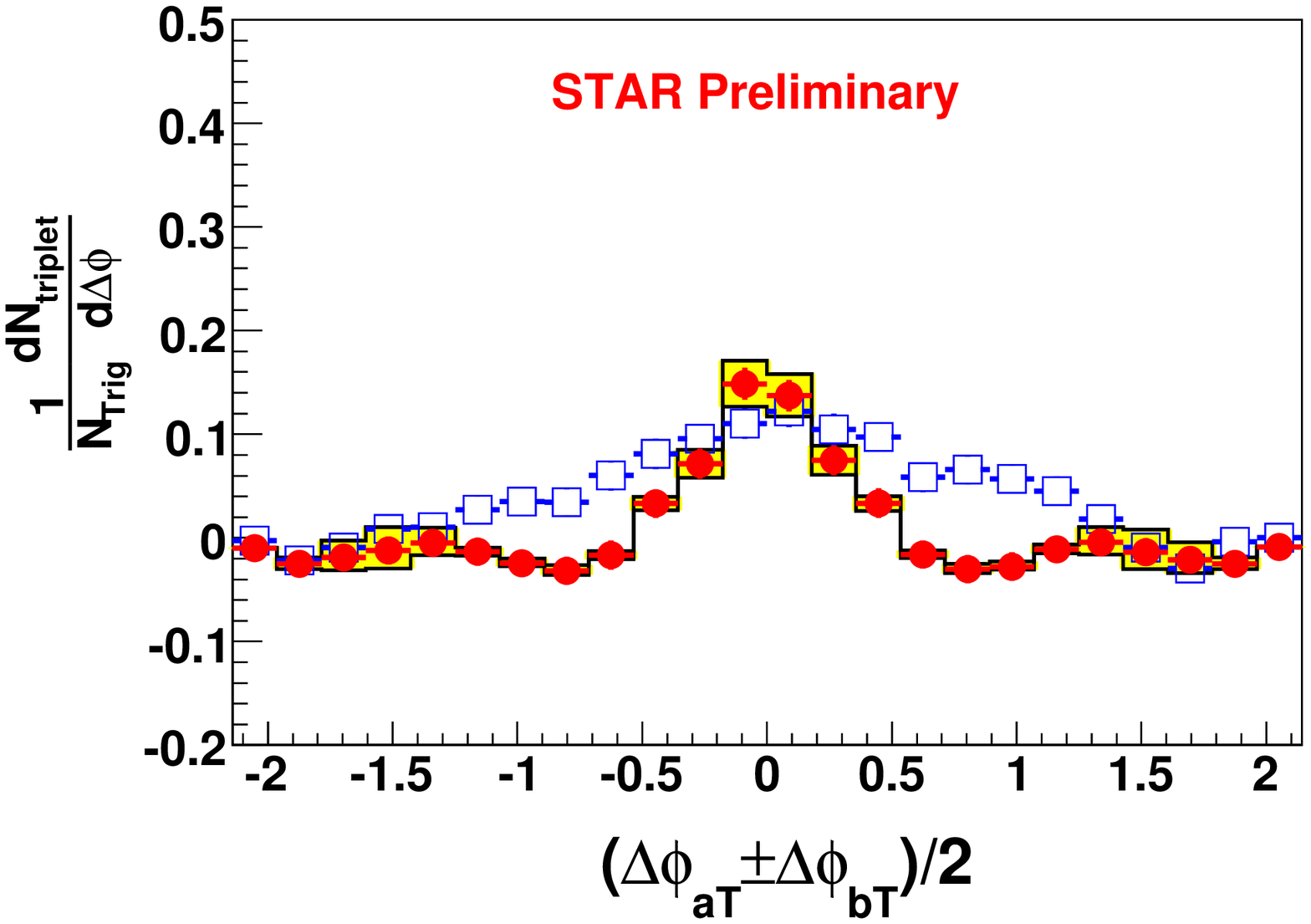} \end{center}\end{minipage}
\begin{minipage}{0.33\linewidth}\begin{center}\includegraphics[width=1.8in,height=2.in]{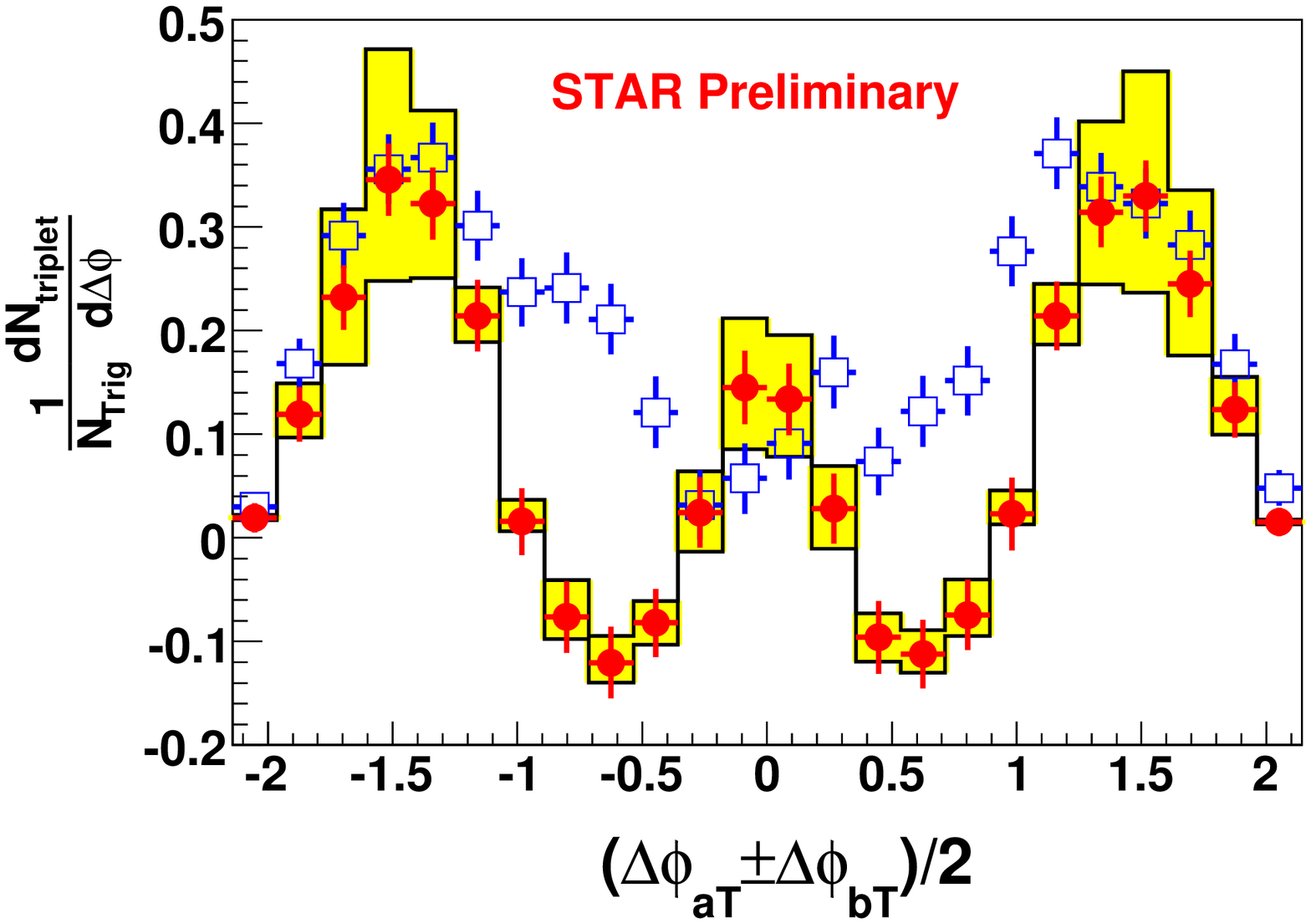} \end{center}\end{minipage}
}
\caption[]{Correlations obtained with model-based subtraction for (left) d+Au, (center) 80-50\% Au+Au, and (right) 12-0\% Au+Au collision. The bottom plots show 
projections along the main (blue) and alternate (red) diagonals.}
\label{fig2}
\end{figure}

\section{Summary}
We presented preliminary results from two on-going STAR searches for conical emission in Au+Au collisions at $\sqrt{s_{NN}}=200$ GeV. 
The cumulant analysis reveals finite three particle correlations dominated by near and away side particle 
correlations consistent with jet production. No clear evidence for conical emission is observed with this method, but the signal may be masked by the presence of
irreducible flow components \cite{Pruneau06}. 
The model-based subtraction method yields similar results for peripheral collisions and reveals structures at ($\pi \pm 1.45$, $\pi \mp 1.45$) suggestive of conical emission in more central collisions.
Further work is in progress to understand the methods sensitivity, systematic effects,  and overall robustness of these results.

{\bf Acknowledgments}
This work was supported, in part, by U.S. DOE Grant No. DE-FG02-92ER40713.

\end{document}